\documentclass[twocolumn, nofootinbib, showpacs]{revtex4-1}

\usepackage{amssymb}
\usepackage{amsmath}
\usepackage{amssymb}
\usepackage{bbold}
\usepackage{amsfonts}
\usepackage{dsfont}
\usepackage[usenames,dvipsnames]{xcolor}
\usepackage{bm}
\usepackage[pdftex]{hyperref}
\hypersetup{plainpages=false,colorlinks=true,linkcolor=blue, citecolor=blue, urlcolor=blue}

\usepackage[english]{babel}
\usepackage{xcolor}
\usepackage{bbold}
\usepackage{mathrsfs}
\usepackage{natbib}

\usepackage{graphicx}
\usepackage{xcolor}

\begin{document}
\title{Single-particle spectral function \\ formulated and calculated by variational Monte Carlo method \\ with application to $d$-wave superconducting state}
\author{Maxime Charlebois$^{1}$}
\thanks{Current affiliation: Center for Computational Quantum Physics, Flatiron Institute, Simons Foundation, 162 5th Ave., New York, 10010 NY, USA}
\author{Masatoshi Imada$^{1,2}$}
\affiliation{$^1$ Waseda Research Institute for Science and Engineering, Waseda University, 3-4-1, Okubo, Shinjuku, Tokyo 169-8555, Japan \\
$^2$ Toyota RIKEN, 41-1, Yokomichi, Nagakute, Aichi 480-1118, Japan }
\date{\today}

\begin{abstract}
A method to calculate the one-body Green's function for ground states of correlated electron materials is formulated by extending the variational Monte Carlo method. We benchmark against the exact diagonalization (ED) for the one- and two-dimensional Hubbard models of 16 site lattices, which proves high accuracy of the method. The application of the method to larger-sized Hubbard model on the square lattice correctly reproduces the Mott insulating behavior at half filling and gap structures of $d$-wave superconducting state  of the hole doped Hubbard model in the ground state optimized by enforcing the charge uniformity, 
evidencing a wide applicability to strongly correlated electron systems.  From the obtained $d$-wave superconducting gap of the charge uniform state, we find that the gap amplitude at the antinodal point is several times larger than the experimental value, when we employ a realistic parameter as a model of the cuprate superconductors.    
The effective attractive interaction of carriers in the $d$-wave superconducting state inferred for an optimized state of the Hubbard model  is as large as the order of the nearest-neighbor transfer, which is far beyond the former expectation in the cuprates.  We discuss the nature of the superconducting state of the Hubbard model in terms of the overestimate of the gap and the attractive interaction in comparison to the cuprates.  
\end{abstract}
\maketitle

\section{Introduction}
Dynamical properties often provide us with crucial insights into open issues of strongly correlated electron systems. 
In particular, momentum and energy resolved single-particle spectral function $A(\mathbf k,\omega)$, which is the imaginary part of the Green's function $G(\mathbf k,\omega)$ with the momentum $\mathbf k$ and the frequency $\omega$ gives us understanding how an electron moves in an environment of other mutually interacting electrons and provides us with properties of the excited states, which in turn reveals the equilibrium properties as well. 

In an example of the copper oxide high-$T_{\rm c}$ superconductors, $A(\mathbf k,\omega)$ has been extensively studied by the angle-resolved photoemission spectroscopy (ARPES), which has greatly contributed to elucidate the properties of superconducting as well as anomalous normal metallic properties including the pseudogap, Fermi arc and $d$-wave superconducting gap structure itself~\cite{damascelli_angle-resolved_2003,lu_angle-resolved_2012}.  

Numerical methods to clarify the dynamics of the strongly correlated electron systems have been hampered by various difficulties such as fermion sign problem~\cite{imada_numerical_1989,loh_sign_1990,furukawa_optimization_1991,troyer_computational_2005} in quantum Monte Carlo methods, and intrinsic quantum entanglement of electrons at long distances. Nevertheless, linear response quantities such as the spin and charge dynamical structure factors, $S(\mathbf k,\omega)$, $N(\mathbf k,\omega)$, respectively, defined below have been studied by limited methods such as the exact diagonalization (ED)~\cite{dagotto_correlated_1994,lanczos_iteration_1950} and time-dependent density matrix renormalization group~\cite{vidal_efficient_2004,white_real-time_2004}. 
However, these methods have their own limitations, namely amenable only in small system sizes and in one-dimensional lattice structure, respectively. 

In addition to the exact diagonalization, single-particle Green's function $G(\mathbf k,\omega)$ has been studied by the cluster extension of the dynamical mean-field theory (cDMFT)~\cite{maier_non-crossing_2000,lichtenstein_antiferromagnetism_2000,kotliar_cellular_2001}, but the allowed momentum resolution is severely limited by the cluster size at most in the order of 10. 
The cDMFT method is generally combined with a periodization procedure in order to restore translation invariance and provides us with the data at interpolated  momenta $\mathbf k$ ~\cite{stanescu_fermi_2006,biroli_cluster_2004,stanescu_cellular_2006,sakai_cluster-size_2012}. However, we need to be cautious about these periodizations and the results should be regarded as estimators, because the momentum resolution is limited by the cluster size anyhow.
This is true even for inhomogenous extension of cDMFT~\cite{charlebois_impurity-induced_2015,faye_interplay_2017}, where large supercluster still retains the self-energy modulation of the original smallest cluster~\cite{verret_intrinsic_2019}. The need to study bigger cluster remains. 

Recent formulation of time-dependent variational Monte Carlo method based on the variational principle opened a way to study the long-time dynamics~\cite{carleo_light-cone_2014,ido_time-dependent_2015}, but it has not been extensively applied yet to interacting fermion systems except for few examples~\cite{ido_correlation-induced_2017}.
Meanwhile, methods of calculating the spin and charge dynamical structure factors utilizing the variational wave functions for ground and excited states have been proposed recently~\cite{li_variational_2010,dalla_piazza_fractional_2015,ferrari_spectral_2018,ferrari_dynamical_2018,ido_charge_2019}. Some attempts to calculate excitation spectrum on larger cluster of the $t$-$J$ model~\cite{yunoki_electron_2007,tan_two-mode_2008}.

Here, we formulate a method of calculating the Green's function $G(\mathbf k,\omega)$ and the spectral function $A(\mathbf k,\omega)=-\frac{1}{\pi}{\rm Im} G(\mathbf k,\omega)$ 
and show its accuracy by comparing with the exact results. It reproduces the feature of the spin-charge separation and excitation continuum in the one-dimensional Hubbard model.

The method is applied to the Hubbard model on the square lattice as well. In the optimized $d$-wave superconducting solution, though it is an excited state in the competition with the stripe states~\cite{darmawan_stripe_2018,ido_competition_2018,zheng_stripe_2017}, the $d$-wave symmetry of the gap structure is correctly reproduced in the spectral function in this charge-uniform lowest energy state. However, the gap amplitude is several times larger than the size in the experimentally observed gap of the cuprate superconductors, if employ a widely accepted parameter mapping. The effective attractive interaction of carriers in this state is then estimated again to be extremely large in the order of or even larger than the nearest neighbor transfer energy $t$ in the Hubbard model.  
We discuss implications of the results.

Finally in the supplementary information the reader can access to a fully functional open source code that was used to generate the data. The code is an extension based on the open source code mVMC~\cite{misawa_mvmcopen-source_2019}.

\section{Method}

\subsection{Green's function}
We present here a very general scheme to estimate one-body Green's function from the Lehman representation.
\begin{align} 
G_{\sigma}(\mathbf k,\omega) &= G^h_{\sigma}(\mathbf k,\omega) + G^e_{\sigma}(\mathbf k,\omega)
\label{GLehman}
\\
G^h_{\sigma}(\mathbf k,\omega) &= \langle\Omega\vert \hat c_{\mathbf k\sigma}^\dagger \frac{1}{\omega+i\eta-\Omega+\hat H} \hat c_{\mathbf k\sigma}\vert \Omega\rangle.
\label{GLehman_h}
\\
G^e_{\sigma}(\mathbf k,\omega) &= \langle\Omega\vert \hat c_{\mathbf k\sigma} \frac{1}{\omega+i\eta+\Omega-\hat H} \hat c_{\mathbf k\sigma}^{\dagger}\vert \Omega\rangle
\label{GLehman_e}
\end{align}
Here $\hat c_{\mathbf k \sigma}$ ($\hat c_{\mathbf k \sigma}^{\dagger}$) annihilates (creates) an electron of momentum $\mathbf k$ and spin $\sigma$.
This approach requires the knowledge of the ground state $\vert\Omega\rangle$ of an Hamiltonian $\hat H$. The ``hat'' notation is used here to represent an operator as opposed to any matrix representation. 
Here $\eta$, is a small real number, a Lorentzian broadening factor.

In the ED, the Lehman representation can be evaluated explicitly since we have a complete and exact representation of the Hamiltonian eigenstates, but it is amenable only to small clusters. To evaluate this Green's function for the cases not amenable to the exact diagonalization, we can use a method similar to the approach used to calculate the spin and charge dynamical structure factors by exhausting important subspace of the Hilbert space for the excitations~\cite{li_variational_2010,dalla_piazza_fractional_2015,ferrari_spectral_2018,ferrari_dynamical_2018,ido_charge_2019}. Namely, the idea of the present method is to restrict Hilbert space of the one particle or hole excited sector of the Hamiltonian to a set of vectors:
\begin{align} 
\vert h_{\mathbf kn}\rangle &= \hat A_n  \hat c_{\mathbf k} \hat B_n \vert \Omega\rangle &&\text{for hole excitations,}
\\
\vert e_{\mathbf kn}\rangle &= \hat A^\dag_n  \hat c^\dag_{\mathbf k} \hat B^\dag_n \vert \Omega\rangle &&\text{for electron excitations,}
\end{align}
where $\hat A_n$ and $\hat B_n$ are operators that together conserve the number of electrons $N_e$ and momentum $\mathbf k$. Here $|\Omega\rangle$ is an approximate ground state of the $N$ particle sector obtained by our variational Monte Carlo method for the ground state. Note that for the Krylov basis of excitation, used in ED, $\hat B_n=\hat I$ and $\hat A_n = \hat H^n$ where $n$ is the number of band Lanczos iteration.  Usually in ED, at every iterations of the band Lanczos method, the excited states basis is orthogonalized to every other excitations. But it is possible and sometime more convenient to work in the non-orthogonal basis. 

\subsection{Non-orthogonal basis}
In the most general case, the excitated states chosen in our method given below are not orthogonal to one another. Then we introduce a number of overlap matrices:
\begin{align} 
O^{h}_{\mathbf k,mn} &= \langle h_{\mathbf km}\vert \hat I \vert h_{\mathbf kn}\rangle 
&
O^{e}_{\mathbf k,mn} &= \langle e_{\mathbf km}\vert \hat I \vert e_{\mathbf kn}\rangle 
\label{Smat}
\end{align}
where $\hat I$ is the identity operator and the matrix notation is expressed with the indices $m$ and $n$. It is important to distinguish the operator notation (with a ``hat'' here) from the matrix notation since they are different in an non-orthogonal basis. Indeed, the matrices $O_{\mathbf k,mn}$ are representations of the identity operator in this non-orthogonal basis. 

Using the same basis, we evaluate the effective Hamiltonian matrices:
\begin{align} 
H^{h}_{\mathbf k,mn} &= \langle h_{\mathbf km}\vert \hat H \vert h_{\mathbf kn}\rangle 
&
H^{e}_{\mathbf k,mn} &= \langle e_{\mathbf km}\vert \hat H \vert e_{\mathbf kn}\rangle 
\label{Hmat}
\end{align}

In general, the basis set on the restricted
Hilbert subspace is nonorthogonal and thus we need to solve the generalized eigenvalue problem within this subspace as:
\begin{align} 
\mathbf H_{\mathbf k} \vert E_{\mathbf k \ell} \rangle  =  
E_{\mathbf k \ell} \mathbf O_{\mathbf k} \vert E_{\mathbf k \ell} \rangle,
\label{Heig}
\end{align}
where $\mathbf H_{\mathbf k}$ and $\mathbf O_{\mathbf k}$ are matrices
whose $(m,n)$ components are given in Eqs. (\ref{Hmat}) and (\ref{Smat}), respectively in the basis of $\vert h_{\mathbf kn}\rangle$ or $\vert e_{\mathbf kn}\rangle$.  
The solution of this generalized eigenvalue problem is represented by the $\ell$-th eigenvalue $E_{\mathbf k \ell}$ and its corresponding eigenstate
coefficients of its eigenvector $\vert E_{\mathbf k \ell} \rangle$ represented in the basis of  $\vert h_{\mathbf kn}\rangle$ or $\vert e_{\mathbf kn}\rangle$. Note that the orthogonality of the eigenvector $\vert E_{\mathbf k \ell} \rangle$ is represented by $\langle  E_{\mathbf k \ell} \vert \mathbf O_{\mathbf k} \vert E_{\mathbf k j} \rangle=\delta_{\ell, j}$.  The eigenvectors expand the basis of the subspace defined by the restricted Hilbert space determined by the choice of non-orthogonal excitations. We can insert the complete set of this subspace $\sum_l \vert E_{\mathbf k l} \rangle \langle E_{\mathbf k l} \vert$ in both Eqs.~\eqref{GLehman_h} and~\eqref{GLehman_e} to obtain:
\begin{align}
G^h_\sigma (\mathbf k,\omega)
&= \sum_{l} \frac{
\langle \Omega \vert
\hat c^\dagger_{\mathbf k\sigma}
\vert E^h_{\mathbf k l} \rangle
\langle E^h_{\mathbf k l} \vert
\hat c_{\mathbf k\sigma}
\vert \Omega \rangle
}{\omega+i\eta - \Omega + E^h_{\mathbf k l}}  
\label{Gh}
\\
G^e_\sigma (\mathbf k,\omega)
&= \sum_{l} \frac{
\langle \Omega \vert
\hat c_{\mathbf k\sigma}
\vert E^e_{\mathbf k l} \rangle
\langle E^e_{\mathbf k l} \vert
\hat c^\dagger_{\mathbf k\sigma}
\vert \Omega \rangle
}{\omega+i\eta + \Omega - E^e_{\mathbf k l}}
\label{Ge}
\end{align}

Note that $|E^{e}_{\mathbf k l}\rangle$ ($ |E^{h}_{\mathbf k l}\rangle$) is the state in the $N+1$ ($N-1$) particle sector with momentum $\mathbf k$. It is important to keep in mind that this is an approximation to the Green's function as we restricted the  $N+1$ or $N-1$ particle sector by a variational form defined below in addition to the approximation to the ground state. By increasing the dimension of the excited subspace, we are able to systematically improve the representation of the excited states toward the exact one represented by the full Hilbert space. The accuracy can be tested by the convergence when we increase the number of variational states. 

Note that we can adapt this formalism to re-express the band Lanczos algorithm.
The main difference here is that we impose the translational symmetry from the beginning and we do not orthogonalize at each iteration aside from the variational form for the ground state. In fact, this is not an iterative technique, but for a chosen Krylov subspace, this formalism would give the same result as the band Lanczos method.

Our goal is to apply this formalism to obtain the spectral function of the strongly correlated fermion systems by combining with the variational Monte Carlo (VMC) method.
An alternative derivation of Eqs.~\eqref{Gh} and~\eqref{Ge} can be found in Appendix~\ref{alternative}. This is the one implemented in the software provided in supplementary information.

\subsection{VMC}

In the variational Monte Carlo, we postulate an ansatz, a variational state $\vert \psi\rangle$ that can be used to calculate the physical quantities associated with that state. To find a good approximation to the ground state, we optimize the variational state in order to minimize the energy measured~\cite{tahara_variational_2008,misawa_mvmcopen-source_2019}. 

The measurement of any operator $\hat A$ can be done as:
\begin{align}
\langle \hat A\rangle&=\frac{\langle\psi|\hat A| \psi\rangle}{\langle\psi | \psi\rangle}=\sum_{x} \frac{\langle\psi|\hat A| x\rangle\langle x | \psi\rangle}{\langle\psi | \psi\rangle}
\label{vmcSummary}
\\
&=\sum_{x} \rho(x) \frac{\langle\psi|\hat A| x\rangle}{\langle\psi | x\rangle}
\;\; \text{where} \;\;  \rho(x)=\frac{|\langle x | \psi\rangle|^{2}}{\langle\psi | \psi\rangle}
\end{align}
In this equation, the sum $\sum_{x} |x\rangle\langle x|$ is a complete set of every possible electronic configuration in the system. For large systems, however, it becomes computationally unfeasible to sum every configuration. Nevertheless we can still estimate this sum using a 
Monte Carlo sampling. The real-space configurations $\{x_{\mathbf{s}}\}$ are generated with the probability $\rho(x_{\mathbf{s}})$. Then,
\begin{align}
\langle \hat A\rangle
&\sim\frac{1}{N_{\text{MC}}} \sum_{x_{\mathbf{s}}} \frac{\langle\psi|\hat A| x_{\mathbf{s}} \rangle}{\langle\psi | x_{\mathbf{s}} \rangle}
\end{align}
where $N_{\text{MC}}$ is the number of Monte Carlo sampling for the summation over $x_\mathbf{s}$, where $\mathbf{s}$ is the index to specify the particle configuration in the real space.


For the ground state wavefunction, we employ the variational state:
\begin{align}
|\psi\rangle&=\mathcal{P}_{\mathcal{G}} \mathcal{P}_{\mathcal{J}} 
|\phi\rangle 
\\
|\phi\rangle&=\left(\sum_{i, j} f_{i j} \hat c_{i \uparrow}^{\dagger} \hat c_{j \downarrow}^{\dagger}\right)^{N_{e} / 2}|0\rangle 
\\ 
\mathcal{P}_{\mathcal{G}}&=\exp \left(\sum_{i} g_i \hat n_{i \uparrow} \hat n_{i \downarrow}\right) 
\\ 
\mathcal{P}_{\mathcal{J}}&=\exp \left(\sum_{i\ne j} v_{i j} \hat n_{i} \hat n_{j}\right)
\end{align}
where the variational parameter are $f_{i j}$, $g_i$ and $v_{i j}$. We define the variational ground state $|\Omega\rangle$ as the state $|\psi\rangle$ that minimizes the variational energy $E = \langle\psi|\mathcal{H}| \psi\rangle / \langle\psi| \psi\rangle$. To simultaneously optimize the variational parameters, the natural gradient method is applied.~\cite{amari_natural_1998,sorella_generalized_2001}.

\subsection{Dynamical VMC}

\begin{figure}
\centering
\includegraphics[width=\linewidth]{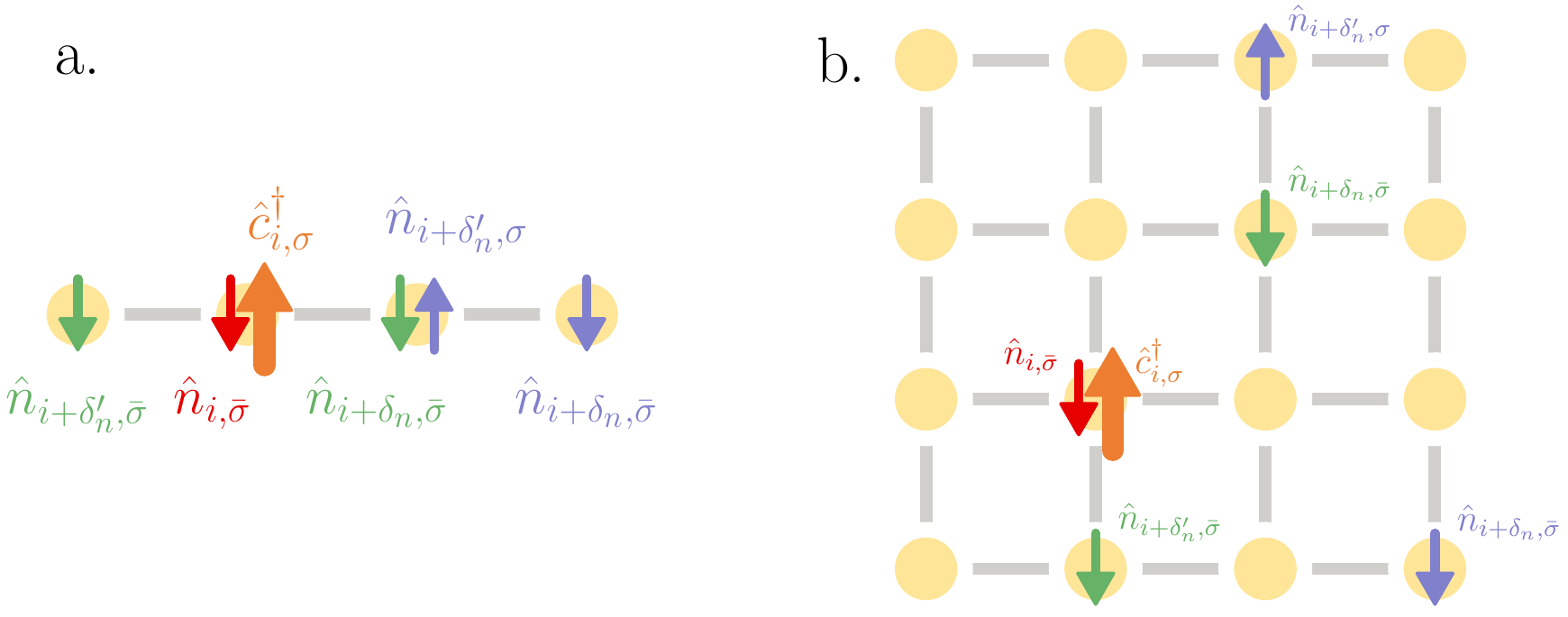}
\caption{Geometries of the system studied in this paper: a. is the one-dimensional lattice and b. is the two-dimensional square lattice. The excitation on site $i$ is correlated with the presence of electron on neighboring sites. The charge of the excitation of type $\hat c^{\dagger}_{i \sigma} \hat n_{i \bar \sigma} \vert \Omega \rangle$ is represented by the red arrow. The charges of the excitation of type $\hat c^{\dagger}_{i \sigma} \hat n_{i+\delta, \bar\sigma} \hat n_{i+\delta^\prime, \sigma} \vert \Omega \rangle$ is represented by the two blue arrows and the charges of the excitation of type $\hat c^{\dagger}_{i \sigma} \hat n_{i+\delta, \bar\sigma} \hat n_{i+\delta^\prime, \sigma} \vert \Omega \rangle$ is represented by the two green arrows.}
\label{1d_chain}
\end{figure}

Now, the excited states are constructed by applying the VMC calculation, because it is hard to generate excited states that reproduce the Krylov basis since the calculation of $\hat H^n$ is very expensive for $n>1$ as every hopping of electron terms in the Hamiltonian produces a new Pfaffian evaluation. Instead we choose a basis where  $\hat A_n=\hat I$ and $\hat B_n$ is a combination of different charge excitations $\hat n_{i\sigma}$\cite{ido_charge_2019}. For example, the excitation basis we choose are
$\vert h_{i\sigma n}\rangle =
\hat c_{i \sigma} \vert \psi_{in} \rangle$ and
$\vert e_{i\sigma n}\rangle = 
\hat c^\dagger_{i \sigma} \vert \psi_{in} \rangle$
where:
\begin{align}
\vert \psi_{in} \rangle =&\{ 
\vert \Omega \rangle,  
\hat n_{i \bar \sigma} \vert \Omega \rangle,  
\nonumber
\\
&
\hat n_{i+\delta_n, \bar\sigma} \hat n_{i+\delta_n^\prime, \sigma} \vert \Omega \rangle,
\hat n_{i+\delta_n, \bar\sigma} \hat n_{i+\delta_n^\prime, \bar \sigma} \vert \Omega \rangle 
\}.
\label{charge_exc}
\end{align}
Here $\delta_n$ and $\delta_n^\prime$ are a combination of different neighbors to site $i$ for each $n$. This choice is based on the physical reason, where the creation of an electron on site $i$ is influenced by the presence of electron with both same and opposite spin on sites $i+\delta$ and $i+\delta^\prime$ respectively. This is illustrated in Fig.~\ref{1d_chain} for the two geometries studied in this article. Generally, we consider only excitation within a certain range $\delta=(\delta_x,\delta_y)$ of neighborhood of the considered site $i$, noted  as $\max(|\delta_{x}|,|\delta_{y}|) \le \delta_{max}$ where $\delta_{max}$ is an integer that specifies the farthest neighbor considered in any direction. It is implicitly applied to both $\delta$ and $\delta^\prime$. Under that threshold, we consider every combination of $\delta$ and $\delta^\prime$ that generate unique new excitation.  The excitations have to be non-redundant (different).  Otherwise this will lead to singular matrices for $\mathbf H_{\mathbf k}$ and $\mathbf O_{\mathbf k}$. Note that $|e_{i\sigma n}\rangle$ is in the electron-number sector with one-electron added to the ground state, while $|h_{i\sigma n}\rangle$ is in the one-electron removed sector.

Equation (\ref{charge_exc}) is the simplest choice of basis to reasonably represent the essential part of low-energy excitation subspace of the Hamiltonian (as demonstrated in the result section). We calculate the average of 
$\langle \psi_{im} \vert \hat c_{i\sigma} \hat c^{\dagger}_{j\sigma} \vert \psi_{jn} \rangle$
and 
$\langle \psi_{im} \vert \hat c_{i\sigma} \hat H \hat c^{\dagger}_{j\sigma} \vert \psi_{jn} \rangle$ 
in the real space representation
for every $i,j$ combination ($N^2$ terms) after Monte Carlo sampling of the Markov chain and then Fourier transform (thanks to the translational symmetry):
\begin{align}
O^{e}_{\mathbf k,mn}
&=
\frac{1}{N_s}\sum_{ij}
e^{-i \mathbf k(\mathbf r_i-\mathbf r_j)}
\langle \psi_{im} \vert \hat c_{i\sigma}  \hat c^{\dagger}_{j\sigma} \vert \psi_{jn} \rangle
\\
H^{e}_{\mathbf k,mn}
&=
\frac{1}{N_s}\sum_{ij}
e^{-i \mathbf k(\mathbf r_i-\mathbf r_j)}
\langle \psi_{im} \vert \hat c_{i\sigma} \hat H \hat c^{\dagger}_{j\sigma} \vert \psi_{jn} \rangle 
\end{align}
This produces a number of matrices as many as twice of the number of the momentum points matrices.
We proceed with the same approach to evaluate the hole counterpart as well  ($O^{h}_{\mathbf k,mn}$ and $H^{h}_{\mathbf k,mn}$) and solve the generalized eigenvalue problem (Eq.~\eqref{Heig}) for both holes and electrons. 

Together with Eqs.~\eqref{GLehman},~\eqref{Gh} and~\eqref{Ge}, we estimate the Green's function from the VMC. We call the technique presented in this section dynamical VMC (dVMC). 
Appendix~\ref{appendix_anticomm} discusses about details to optimize the computation speed of the calculation of these excitations.

\section{Results}
\subsection{Model}

\begin{figure}
\centering
\includegraphics[width=\linewidth]{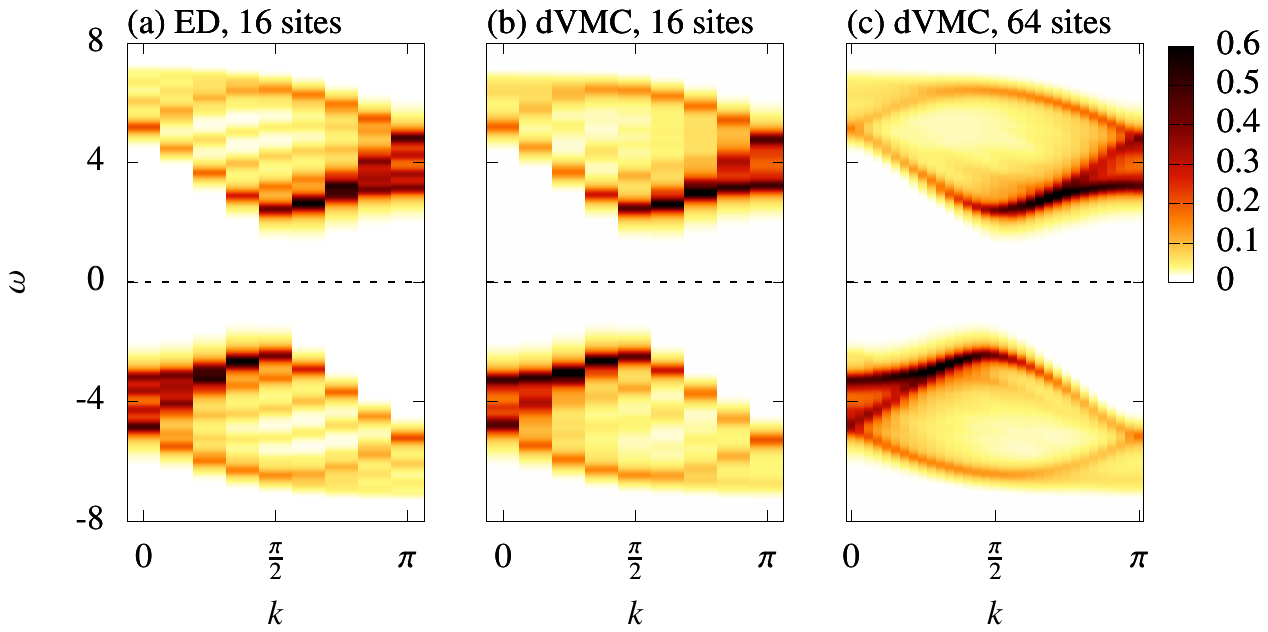}
\caption{Comparison of spectral weight for the 1D Hubbard model at half filling ($L=16$ and $N_e=16$) calculated by the ED (a) and by the present method dVMC (b). For the 16-site dVMC calculation, we considered combinations of every possible neighbors for the choice of $\delta$ and $\delta^\prime$ in Eq.~\eqref{charge_exc}. Since the size is small, $N_{exc}=377$ excitations are considered in total by omitting redundant excitations. In (c) we show the result for dVMC with $64$ sites. We considered up to $\delta_{max} = 8$, resulting in a total of $N_{exc}=426$ excitations. }
\label{chain16_U8_Ne16}
\end{figure}

\begin{figure}
\centering
\includegraphics[width=\linewidth]{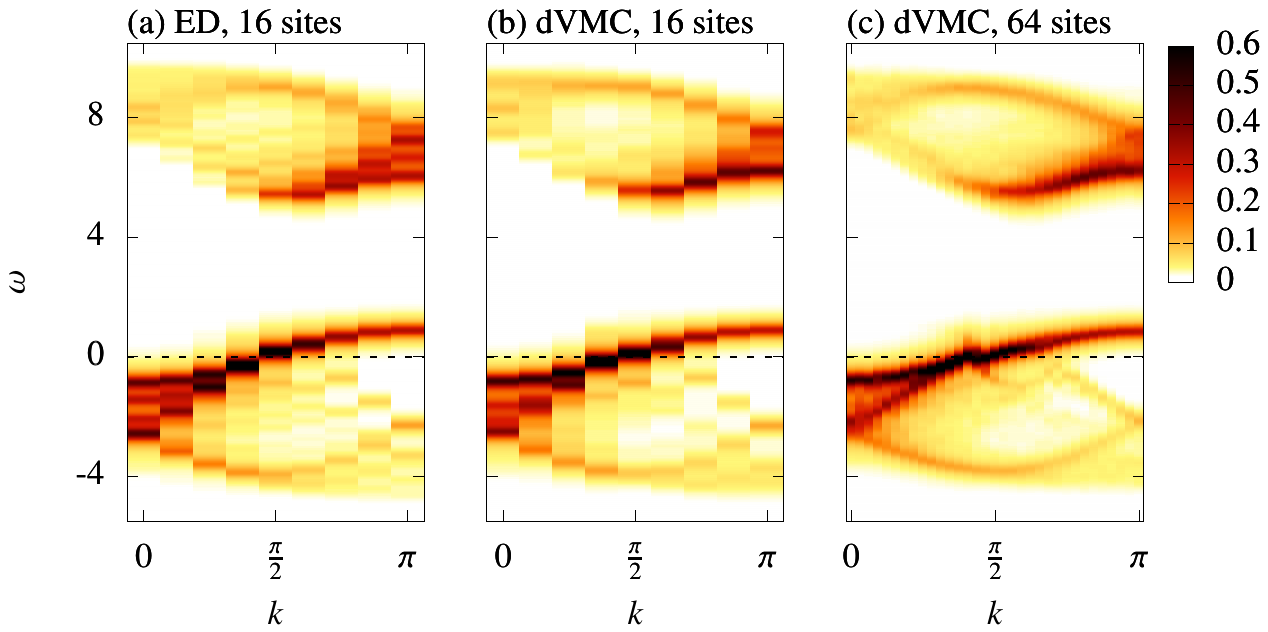}
\caption{Comparison of spectral weight for the doped 1D Hubbard model
at $U/t=8$ ($L=16$ and $N_e=14$) (b) with the ED result (a). Results of lightly doped 64-site chain ($N=64$ and $N_e=56$) is shown in (c).  Here, $\omega=0$ is the Fermi level. The choices of $\delta$ and $\delta'$ for (b) and (c) are the same as in Fig.~\ref{chain16_U8_Ne16}.}
\label{chain16_U8_Ne14}
\end{figure}

To examine the accuracy of the present dynamical VMC method, we show the benchmark test of the standard Hubbard model with the Hamiltonian 
\begin{align}
\hat H=-t \sum_{\langle i, j\rangle, \sigma} \hat c_{i \sigma}^{\dagger} \hat c_{j \sigma}+U \sum_{i} \hat n_{i \uparrow} \hat n_{i \downarrow}
\end{align}
on the 1D chain and 2D square lattice.
The first term proportional to the transfer $t$ is the kinetic energy and the sum is restricted to the nearest neighbor pair. The second term represents the onsite Coulomb repulsion proportional to $U$.
We impose the periodic boundary condition throughout this paper.
In the presentation, the same color scale is used throughout this paper to make the comparison between different methods and size easier, except for Fig.~\ref{fig:gap}. Unless specified, we use parameters $U=8$, and the nearest neighbor transfer $t=1$ as the energy unit.
We add a broadening factor $\eta=0.2$. 
The chemical potential (Fermi level) is determined so as to meet the occupied part of the integrated spectral weight with the number of electrons ($N_e =$ integer) given in our canonical ensemble simulation of VMC. 
Results obtained by the present method on lattices of $N=16$ sites are first compared to ED results in order to benchmark the accuracy of the methods in one and two dimensions. The ED results were obtained using the open-source software H$\Phi$~\cite{kawamura_quantum_2017}

\subsection{One-dimensional lattice}

In this section we test both the Mott insulator ($N=N_e$)  and the doped Mott insulator in one-dimensional Hubbard model by comparing with the exact diagonalization results for the 16-site chain. The results at half filling are shown in Fig.~\ref{chain16_U8_Ne16}(b) in comparison with exact diagonalization results in (a).  We see that the present dVMC results show nearly perfect and quantitative agreement about the Mott insulating nature between the results of the present method and the ED in terms of the dispersion of the lower (occupied) Hubbard band below $E_F$ and the upper (unocccupied) Hubbard band above $E_F$, in terms of their broadnesses, relative weights and the Mott gap sizes.

The results for the 64-site chain is shown in Fig.~\ref{chain16_U8_Ne16}(c), which shows much higher momentum resolution. Both in 16-site and 64-site results, the lower (upper) Hubbard dispersion below (above) $E_F$ has split into two dispersions around the $\Gamma$ ($k=0$) and $k=\pi$ points. The upper flat dispersion branch in the lower Hubbard dispersion was identified as the spinon branch and the lower steeper dispersion branch was identified as coming from the holon excitation in the analysis of the ARPES data for SrCuO$_2$~\cite{kim_observation_1996,kim_separation_1997}. The present dVMC calculation correctly captures the spin charge separation in the 1D Hubbard model. 

The doped case is shown in Fig.~\ref{chain16_U8_Ne14}. The comparison between the dVMC result in (b) with the ED result in (a) for the doping concentration $\delta=0.125$ again shows nearly perfect agreement. The same doping for the 64-site systems in Fig.~\ref{chain16_U8_Ne14}(c) shows good agreement with the result obtained by Kohno using the dDMRG for 60-site system with the open boundary condition~\cite{kohno_spectral_2010,kohno_characteristics_2018}. We still see both the holon and spinon bands. We also note the presence of the hole-pocket behavior appearing at the Fermi level~\cite{kohno_spectral_2010}.
We show the dependence on the number of excitations taken into account
in Appendix~\ref{decomposition_appendix}.

\subsection{Two-dimensional square lattice}

\begin{figure}
\centering
\includegraphics[width=\linewidth]{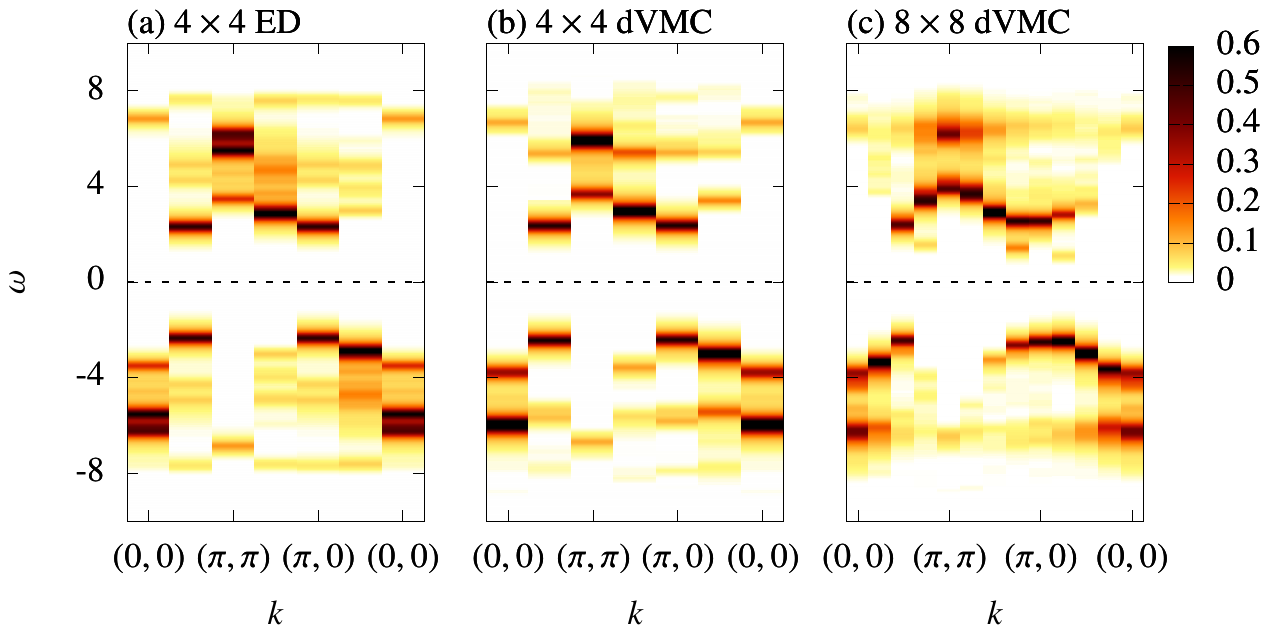}
\caption{Comparison of spectral weight $A(k,\omega)$ in the momentum-energy plane for the square-lattice Hubbard model at half filling ($N=4 \times 4$ and $N_e=16$) calculated by ED (a) and by dVMC (b).  For the 16-site dVMC calculation, we considered combinations of every possible neighbors for the choice of $\delta$ and $\delta^\prime$ in Eq.~\eqref{charge_exc}. In total, $N_{exc}=377$ excitations are taken into account after omitting redundant excitations. The result of the same quantity for the $8\times 8$ lattice is shown in (c),
which shows an essential agreement with the result of the cluster perturbation theory~\cite{kohno_mott_2012,kohno_characteristics_2018}. In all (a), (b) and (c), $A(k,\omega)$ is plotted along the symmetric line of the momentum in the Brillouin zone.}
\label{8x8_Ne64_Akw}
\end{figure}

\begin{figure}
\centering
\includegraphics[width=\linewidth]{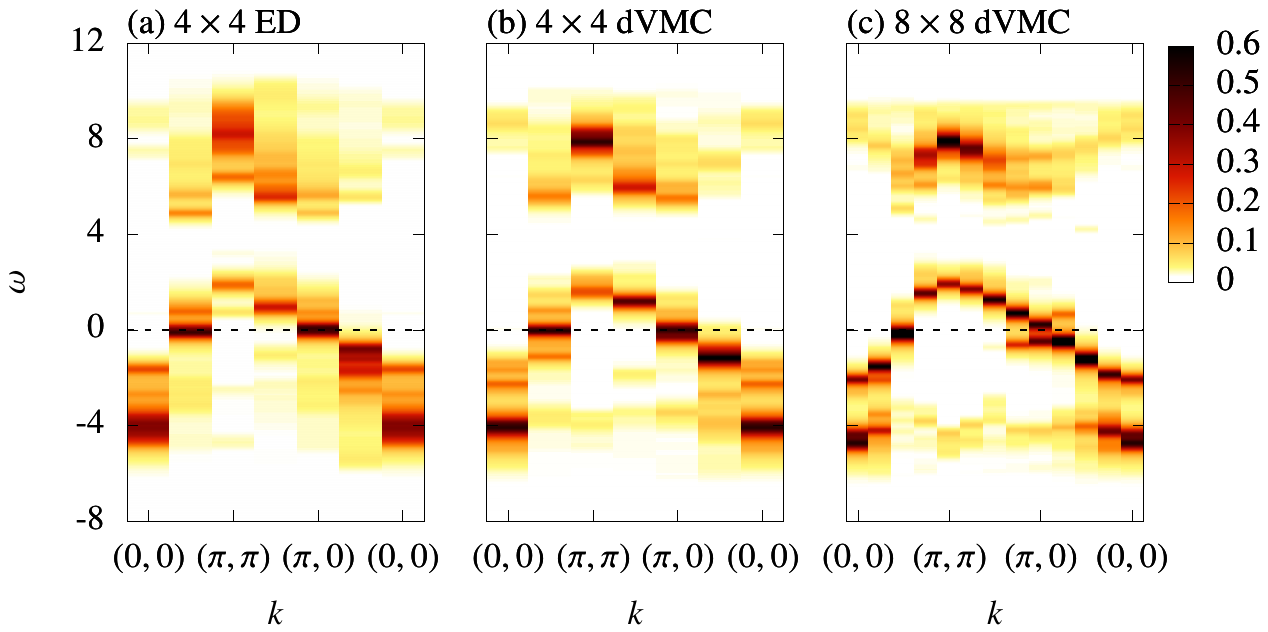}
\caption{Spectral weight for doped square-lattice Hubbard model ($N=4 \times 4$ and $N_e=14$) in (b) compared with the ED result in (a). The result of $8\times 8$ lattice with $N_e=56$ in (c) at 12.5\% hole doping is shown in (c). The momentum dependence is shown along the symmetric line in the same way as Fig.\ref{8x8_Ne64_Akw}.
$\omega=0$ is the Fermi level.}
\label{8x8_Ne56_Akw}
\end{figure}

In this section, we examine the square lattice both for the Mott insulator ($N=N_e$) (see Fig.~\ref{8x8_Ne64_Akw}), and the doped Mott insulator (Fig.~\ref{8x8_Ne56_Akw}) again. Comparison of the dVMC results in Fig.~\ref{8x8_Ne64_Akw}(b) and Fig.~\ref{8x8_Ne56_Akw}(b) with the exact diagonalization results in Fig.~\ref{8x8_Ne64_Akw}(a) and Fig.~\ref{8x8_Ne56_Akw}(a), respectively,  proves good accuracy of the dVMC even for the 2D lattice.  The comparison of $8\times 8$ lattice results in Fig.~\ref{8x8_Ne64_Akw}(c) at half filling and Fig.~\ref{8x8_Ne56_Akw}(c) at 12.5$\%$ hole doping with the result of the cluster perturbation theory~\cite{kohno_mott_2012,kohno_characteristics_2018} shows a fair overall agreement. Note that the results of the cluster perturbation has fewer momentum resolution and should not be taken as a sufficiently accurate reference. Therefore, a small discrepancy does not necessarily mean the insufficiency of the present result.  

In both Fig.~\ref{8x8_Ne64_Akw}(c) and Fig.~\ref{8x8_Ne56_Akw}(c), we have employed a $2\times2$ sublattice to constrain the ground state to exclude stripe orders as well as $x$-$y$ anisotropies~\cite{ido_competition_2018}. 
Fig.~\ref{8x8_Ne56_Akw}(c) shows results for a low-energy state, which has a superconducting order with $d_{x^2-y^2}$ symmetry. Note that the true ground state may have the stripe order with vanishing superconducting order~\cite{ido_competition_2018,darmawan_stripe_2018,zheng_stripe_2017}. However, the superconducting state is very close to the ground state and is obtained at least as a well optimized metastable state.  Therefore, we could expect that we are able to see the intrinsic property of the superconducting phase when it is stabilized in realistic Hamiltonian~\cite{ohgoe_ab_2019}. The superconducting correlation function for this state confirms the superconducting long range order (see Appendix~\ref{Pd}).

\begin{figure}
\centering
\includegraphics[width=\linewidth]{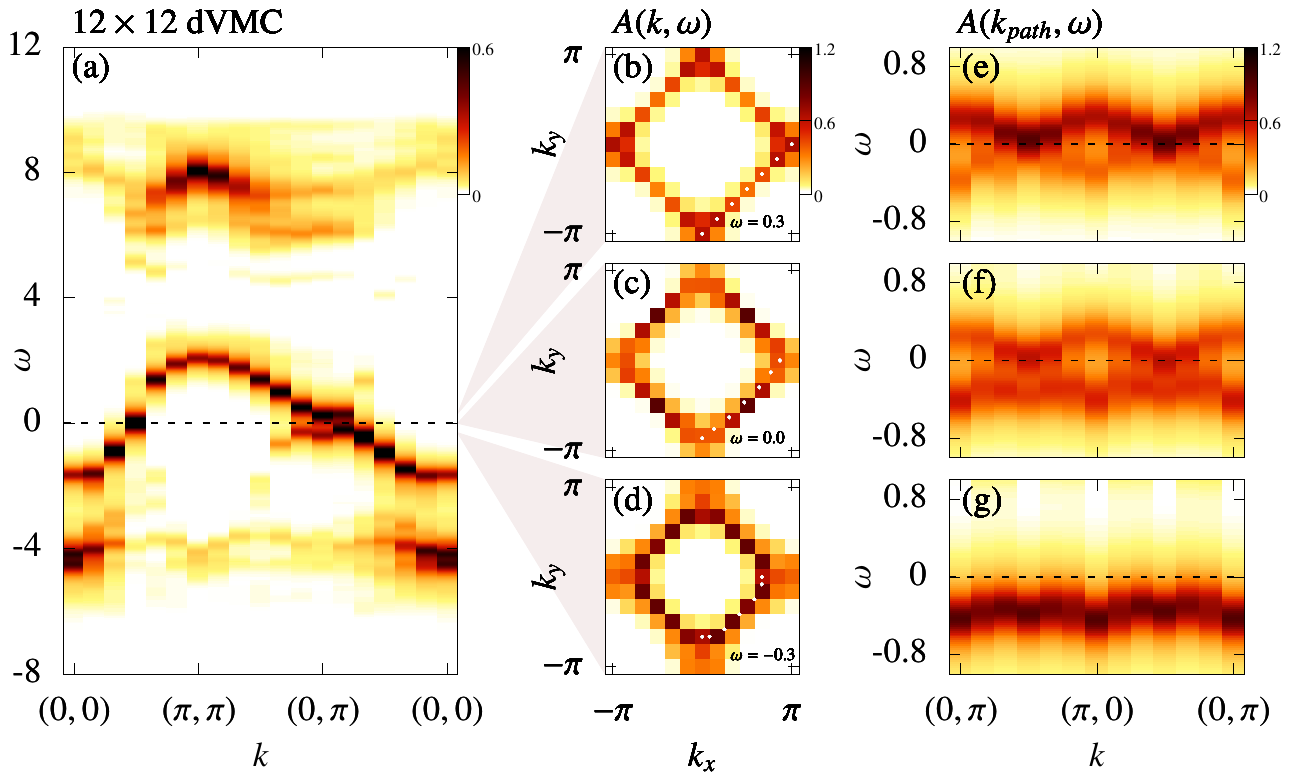}
\caption{Spectral function $A(k,\omega)$ for a $12 \times 12$ cluster. (a) Same view as Fig.~\ref{8x8_Ne56_Akw}(c). (b), (c), (d) Spectral function for a constant energy at and around Fermi level ($\omega=0.0$) plotted in the Brillouin zone. Energies are respectively $\omega=0.3, 0.0, -0.3$. (e), (f), (g) Spectral function in the momentum-energy plane with the momentum following the symmetric line along $(0,\pi)-(\pi,0)-(0,\pi)$. The labels of $\mathbf k$ axis on the abscissa in these panels show only the points on the trajectory of white dots closest to the symmetric points, because the white dots trajectory do not precisely along the symmetric line. We follow the maximum intensity lines in the Brillouin zone of the corresponding panel on the left. The choice of $k$ points in the path shown for panel (e) is shown with the white dots on the corresponding panel on the left (panel (b)). The same is true for panels (c) and (f) and for panels (d) and (g), but in these cases we used linear interpolations to more precisely identify the paths away from the discrete $k$ points. The superconducting $d$-wave gap structure is clearly seen on panel (f).
Note that in this figure, (a) has the same color scale as the rest of the paper, but (b) to (g) has another color scale (shown in top right), in order to see more clearly the details in the superconducting gap and around the Fermi level.}
\label{fig:gap}
\end{figure}

To have a better view of the superconducting physics, we reproduced the results on a bigger cluster. In Fig.~\ref{fig:gap}, we examine a $12\times 12$ cluster with the same parameters as Fig.~\ref{8x8_Ne56_Akw}(c). We see similarities between Fig.~\ref{fig:gap}(a) and Fig.~\ref{8x8_Ne56_Akw}(c) but with much more details. To see the detailed structure of the gap, we first plot, in Fig.~\ref{fig:gap}(b)-(d) the spectral weight in the Brillouin zone for energies at and near the Fermi level, namely  $\omega=-0.3, 0.0$ and $0.3$ to estimate the position of the gap opening (in other words the locus of the minimum gap, which should form the Fermi surface when the gap closes). In principle, the locus of the gap opening can be inferred from the maximum intensity of $A(k,\omega=0)$. The gap anisotropy can be seen along this trajectory. However, because of the limited discrete points in the Brillouin zone allowed for finite size studies, a better estimate of the gap opening position is estimated by an interpolation of $A(k,\omega)$ between neighboring momentum points using the data at small but nonzero $\omega$ as well.   In fact, the rough estimate of the largest gap indicates that it appears at the antinode with the amplitude $\Delta \sim 0.3 t$. Therefore, the interpolation can be made efficiently by the choice of $|\omega|\le 0.3t$. At $\omega/t=-0.3, 0.0$ and $0.3$, we also show thus obtained trajectory of the maximum $A(k,\omega)$ intensity by white dots and the corresponding interpolated spectral weight for these dots on the right, in panels (e), (f), (g). It is important to interpolate between different $\omega$ because the Fermi level determined from the consistency between the electron number in the occupied part of $A(k,\omega)$ and the given nominal number itself has uncertainty arising from the discrete $k$ points and the broadening factor. 
In addition, because of the gap, it contains another uncertainty in identifying precisely the gap opening position in Fig.~\ref{fig:gap}(c).  Nevertheless, in Fig.~\ref{fig:gap}(f), we can see clearly that the gap anisotropy is essentially expressed by the {\it d}-wave superconducting gap ($\Delta(\mathbf k) = \frac{\Delta}{2} (\cos k_x - \cos k_y) $), which has the maximum at the anti-nodes ($\mathbf k = (0,\pm\pi), (\pm\pi,0)$) and closes at the nodes ($k_x = k_y$), though the precise functional form of the gap is beyond the scope of the present paper. 

Note that the number of excitation, for the case in Fig.~\ref{8x8_Ne56_Akw}(c) and Fig.~\ref{fig:gap}, have been limited to 118 excitations, by keeping only the range $0 \le \delta^{(\prime)}_{x} \le 2$ and $0 \le \delta^{(\prime)}_{y} \le 2$ to suppress the statistical error within our allowed computational cost.

The ratio $V_{d}=\Delta / F$ with $F=\sqrt{P_d(r\rightarrow\infty)}$ being the superconducting order parameter $\langle \Delta_d \rangle$ is the measure of the effective attractive interaction to form the Cooper pair. 
From Fig.~\ref{Pd_r} in Appendix~\ref{Pd}, $\langle \Delta_d \rangle$ is estimated to be 0.055.
Then the attractive interaction is roughly estimated as $V_d\sim 1.7t$, which is extremely large, implying that the superconductivity in the Hubbard model is unrealistically strong, if we take the Hubbard model as a model of the cuprate superconductors with $t\sim 0.5$ eV as employed in the literature. In fact, the gap size itself is estimated to be $\sim 0.3t$, interpreted as $\sim 150$ meV for the cuprates, which is several times larger than the gap amplitude around or less than 50 meV observed in the cuprates~\cite{damascelli_angle-resolved_2003,lu_angle-resolved_2012}.    However, as we mentioned above, the true ground state is replaced by the charge inhomogeneous stripe state because of such a strong attraction. 

A view for the origin of the effective attraction is the energy gain by the recovery of the electron coherence (fading out of Mottness), which grows in a nonlinear fashion with evolution of the doping. This nonlinear reduction of the kinetic energy forming a strong convex upward curve of the electronic energy as a function of the carrier density with the negative curvature signals the effective attractive interaction of careers, because the quadratic term obviously represents the effective electron-electron interaction~\cite{imada_excitons_2019}. The attractive interaction estimated from the negative curvature again has the same order consistently with the present value $\sim t$~\cite{misawa_origin_2014}.

By using the present method, we have shown that the $d$-wave gap of the Hubbard model has the energy scale of $t$ (more precisely $\sim 0.3t$), which is much larger than the value as the model of the cuprate superconductors.  We note that more realistic understanding of the scale of the attraction of the cuprates and other available superconductors has to be reached by using the {\it ab initio} effective Hamiltonian that has realistic nonzero off-site Coulomb repulsions~\cite{ohgoe_ab_2019}.

\section{Summary and Outlook}
In this paper we examined a newly proposed dVMC method to calculate the single-particle spectral function and the Green's function for strongly correlated electron systems. 
Although the proposed variational form of the excited states are simple and contains only one bare electron or hole added to the ground state dressed by composite operators diagonal in the particle number representation, the obtained spectral function rather accurately reproduces the exact structure in the benchmark test.

An application to the hole-doped square-lattice Hubbard model revealed that a $d$-wave superconducting state is induced by an effective carrier attraction, which is unexpectedly large in the order of the electron transfer resulting in much larger superconducting gap than that observed in the corresponding cuprates, if we study the charge uniform lowest-energy state as the ground state. It implies that the real cuprate superconductors have to be understood by taking account of more realistic factors such as the intersite Coulomb repulsion, which suppresses both the charge inhomogeneity and the superconducting order overestimated in the Hubbard model in comparison to the real existing superconductors.

Though it reached unprecedented and fruitful results, the obtained spectral function is not perfect with some discrepancy from the exact results. 
We note that spectral functions are very sensitive to the ground state $\vert \Omega \rangle$ used in the calculation, which are severely competing with other metastable states. 
The sensitivity requires a high accuracy of the ground state wave function, before calculating the dVMC Green's calculation. In addition, the form of the excited states has to be flexible enough particularly to represent low-energy excitations. 
Qualitatively different types of excitations ignored in the present work but presumably being important are the dressing by the spin-flip excitation such as $\hat c^\dagger_{i+\delta, \sigma} \hat c_{i+\delta, \bar\sigma} \hat c^\dagger_{i, \bar\sigma} \vert \Omega \rangle$ and by kinetic operators such as $\hat c^\dagger_{i+\delta, \sigma} \hat c_{j+\delta, \sigma}  c^\dagger_{i, \sigma} \vert \Omega \rangle$ which are off-diagonal in the particle number representation. The inclusion of these excitations is an intriguing future issue to enhance the accuracy. Of course increasing the number of charge operators as $\hat n_{i, \bar\sigma} \hat n_{i+\delta_n^\prime, \sigma} \hat n_{i+\delta_n^{\prime\prime}, \bar\sigma} \vert \Omega \rangle$, $\hat n_{i, \bar\sigma} \hat n_{i+\delta_n^\prime, \sigma} \hat n_{i+\delta_n^{\prime\prime}, \sigma} \vert \Omega \rangle$ may also improve the accuracy.

In fact, the present dVMC method with such an improvement is expected to contribute to better understanding of the low-energy subtle structures such as pseudogap and effect of severe competitions among superconducting, charge and spin correlations of the doped Mott insulator.

\section{Acknowledgments}
\label{acknowledgments}

We acknowledge Youhei Yamaji, Kota Ido, Takahiro Ohgoe and Shiro Sakai for fruitful discussions. This work was supported by Fonds de recherche du Qu\'{e}bec - Nature et technologies (FRQNT) and by a Grant-in-Aid for Scientific Research (No. 16H06345) from Ministry of Education, Culture, Sports, Science and Technology, Japan
The authors are grateful to the MEXT HPCI Strategic Programs, and the Creation
of New Functional Devices and High-Performance Materials to Support Next Generation Industries (CDMSI) for their financial support.
We also acknowledge the support provided by the RIKEN
Advanced Institute for Computational Science under the
HPCI System Research project (Grants No. hp180170 and hp190145).
A part of the computation was done at Supercomputer Center, Institute for Solid State Physics, University of Tokyo.

\appendix

\begin{figure*}
	\includegraphics[width=\textwidth,height=4cm]{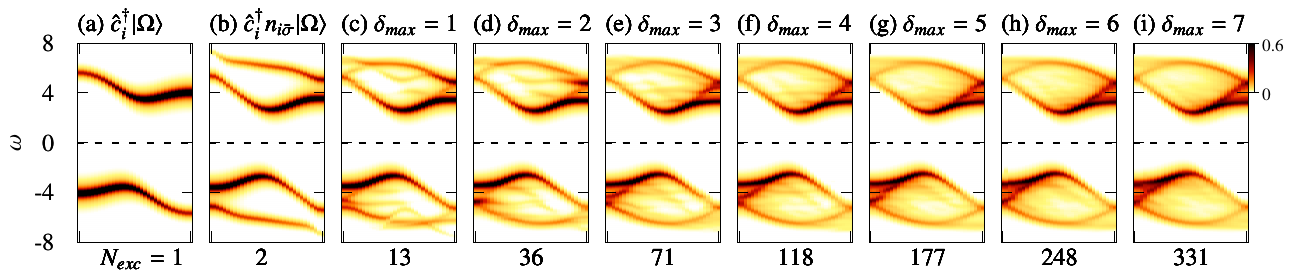}
	\caption{Comparison of spectral weight for the 64-site 1D Hubbard model in Mott insulating state, as shown in Fig.~\ref{chain16_U8_Ne16}(c), but considering only a subset of excitations. Indeed, in Fig.~\ref{chain16_U8_Ne16}(c), we consider  up to a threshold $\delta_{max} = 8$. Here we show the result for $\delta_{max} = 7$ on panel (i), $\delta_{max} = 6$ on panel (h) and so on. Panel (b) plots the case where the first two (local) excitations of Eq.~\eqref{charge_exc} are taken into account and panel (a) is the case taking into account only the first one, {\it i.e.} the trivial excitation. $\mathbf k$ axis are the same as in Fig.~\ref{chain16_U8_Ne16}(c) so they have been omitted here.}
	\label{chain64_decomp_mott}
\end{figure*}

\begin{figure*}
	\includegraphics[width=\textwidth,height=4cm]{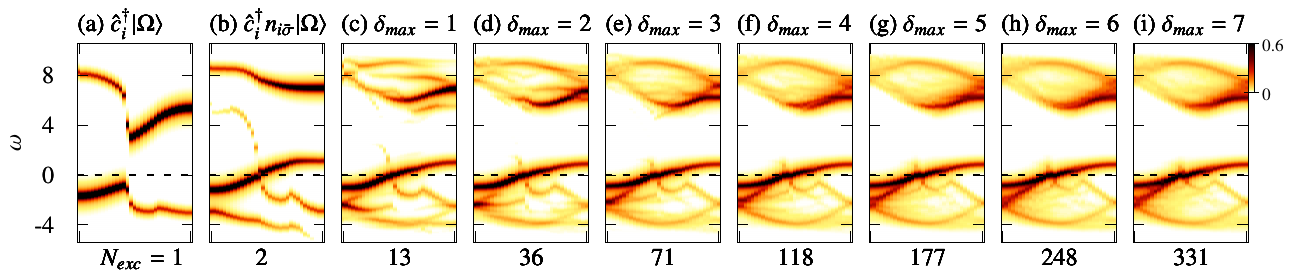}
	\caption{Comparison of spectral weight for the 64-site 1D Hubbard model at $12.5\%$ hole doping, as shown in Fig.~\ref{chain16_U8_Ne14}(c), but considering only a subset of excitation. Details are the same as Fig.~\ref{chain64_decomp_mott}.}
	\label{chain64_decomp}
\end{figure*}

\section{Alternative proof}
\label{alternative}

This appendix shows the Green's function calculation as it is implemented in the dVMC code available in Supplementary Information. It is an alternative form of Eqs~\eqref{Gh} and~\eqref{Ge}.

The Green's function in a non-orthogonal basis is also expressed as~\cite{soriano_theory_2014}:
\begin{align}
\mathbf G_h(\mathbf k,\omega) 
&= \mathbf O_{h\mathbf k} \big((\omega + i \eta - \Omega) \mathbf O_{h\mathbf k} + \mathbf H_{h\mathbf k} \big)^{-1} \mathbf O_{h\mathbf k}
\label{Gmat1}
\\
\mathbf G_e(\mathbf k,\omega) 
&= \mathbf O_{e\mathbf k} \big((\omega + i \eta + \Omega) \mathbf O_{e\mathbf k} - \mathbf H_{e\mathbf k} \big)^{-1} \mathbf O_{e\mathbf k} 
\label{Gmat2}
\end{align}
which are related to the the abstract Green operators $\hat G_h(\mathbf k,\omega)$ and $\hat G_e(\mathbf k,\omega)$ defined as:
\begin{align}
\mathbf G_h(\mathbf k,\omega) \vert_{mn} &= \langle h_{\mathbf km}\vert  \big({(\omega + i \eta - \Omega)\hat I  +\hat H}\big)^{-1} \vert h_{\mathbf kn}\rangle
\nonumber
\\
\mathbf G_e(\mathbf k,\omega)\vert_{mn} &= \langle e_{\mathbf km}\vert  \big({(\omega + i \eta + \Omega)\hat I  -\hat H}\big)^{-1} \vert e_{\mathbf kn}\rangle.
\nonumber
\end{align}
From these definitions, we see that if we choose the first excitations such that $\hat B_0=\hat I$, we obtain the Green's function~\eqref{GLehman} with:
\begin{align}
G_{\mathbf k\sigma}(\omega) &= (\mathbf G_e(\mathbf k,\omega) + \mathbf G_h(\mathbf k,\omega))_{m=n=0}.
\label{Gproof}
\end{align}

To speed up calculations, we can rearrange and diagonalize the terms $\mathbf H_{h\mathbf k} \mathbf O^{-1}_{h\mathbf k}$  and $\mathbf H_{e\mathbf k} \mathbf O^{-1}_{e\mathbf k}$ in Eqs.~\eqref{Gmat1} and ~\eqref{Gmat2}:
\begin{align}
G_{\mathbf k\sigma}(\omega)
&= \sum_{l} \frac{
	(\mathbf U_{h\mathbf k})_{0l}
	(\mathbf U^{-1}_{h\mathbf k}\mathbf O_{h\mathbf k})_{l 0}
}{z - \Omega + E_{h\mathbf k,l}}  
+
\frac{
	(\mathbf U_{e\mathbf k})_{0l}
	(\mathbf U^{-1}_{e\mathbf k}\mathbf O_{e\mathbf k})_{l 0}
}{z + \Omega - E_{e\mathbf k,l}} 
\end{align}
where the energies $E_{\mathbf k,l}$ are the eigenvalues and $\mathbf U_{\mathbf k}$ are the eigenvector matrices. Note that even if both matrices $\mathbf H_{\mathbf k}$ and  $\mathbf O^{-1}_{\mathbf k}$ are hermitian, their product is not. Hence $\mathbf U_{\mathbf k}$ is not unitary. This form is equivalent to Eqs~\eqref{Gh} and~\eqref{Ge}.  

\section{Sampled quantities}
\label{appendix_anticomm}

The bottleneck of the calculation of $O_{\mathbf k,mn}$ and $H_{\mathbf k,mn}$ for both hole and electron is the Monte Carlo sampling with a charge configuration $\vert x_{\mathbf{s}} \rangle$ of the quantities:
\begin{align}
g_{ij,\sigma}(x_{\mathbf{s}}) &= \langle \Omega \vert \hat c^{\dagger}_{i\sigma} \hat c_{j\sigma} \vert x_{\mathbf{s}} \rangle
\label{green1}
\\
g_{ijkl,\sigma\sigma^\prime} (x_{\mathbf{s}}) &= \langle \Omega \vert \hat c^{\dagger}_{i\sigma} \hat c_{j\sigma} \hat c^{\dagger}_{k\sigma} \hat c_{l\sigma} \vert x_{\mathbf{s}} \rangle
\label{green2}
\end{align}
for $0 \le i,j < N $ and $k,l$ are the indices of every hopping on the cluster. This requires then $2N^2 (1+2 N_t)$ Pfaffian evaluations in total (where $N_t$ is the number of hoppings in the cluster) for the simple Hubbard model. When we have all the values of Eqs.~\eqref{green1} and~\eqref{green2} in hand,  from these values alone, we can deduce: 
\begin{align}
\langle \psi_{im} \vert \hat c_{i\sigma} \hat c^{\dagger}_{j\sigma} \vert \psi_{jn} \rangle
&&
\langle \psi_{im} \vert \hat c^{\dagger}_{i\sigma} \hat c_{j\sigma} \vert \psi_{jn} \rangle
\label{terms1}
\\
\langle \psi_{im} \vert \hat c_{i\sigma} \hat H \hat c^{\dagger}_{j\sigma} \vert \psi_{jn} \rangle
&&
\langle \psi_{im} \vert \hat c^{\dagger}_{i\sigma} \hat H \hat c_{j\sigma} \vert \psi_{jn} \rangle
\label{terms2}
\end{align}
straightforwardly by using (anti-)commutation relations of fermion operators for the simple Hubbard model.

First of all, since the inserted state $\vert x_{\mathbf{s}} \rangle$ in the VMC sampling method~\eqref{vmcSummary} is a real space configuration, the evaluation of any $\hat n_{i\sigma}$ is fast since $\hat n_{i\sigma} \vert x_{\mathbf{s}} \rangle = \vert x_{\mathbf{s}} \rangle  n_{i\sigma}(x_{\mathbf{s}})$ where $n_{i\sigma}(x_{\mathbf{s}})$ is a scalar: $1$ if the site $i$ of spin $\sigma$ is occupied, and $0$ otherwise. The set of charge diagonal operators of the left hand side can be moved to the right analytically until they reach the right inserted $\vert x_{\mathbf{s}} \rangle$. For this purpose we use the relation:
\begin{align}
\hat n_{a} 
\bigg(
\prod_i
\hat c^{\dagger}_{i} 
\prod_j
\hat c_{j} 
\bigg)
&=
\bigg(
\prod_i
\hat c^{\dagger}_{i} 
\prod_j
\hat c_{j} 
\bigg)
\big(
\hat n_{a} 
+ \sum_i \delta_{ia}
- \sum_j \delta_{ja}
\big)
\end{align}
to commute any charge operator from left to right (the order of $\hat c^{\dagger}_{i} $ and $\hat c_{j} $ does not matter). The terms in Eq.~\eqref{terms1} and the interaction part ($\hat H_U$) of~Eq.\eqref{terms2} can thus all be computed from the values:
\begin{align}
\langle \Omega \vert \hat c_{i\sigma} \hat c^{\dagger}_{j\sigma} \vert x_{\mathbf{s}} \rangle
&=
\delta_{ij}-g_{ji,\sigma}(x_{\mathbf{s}})
\\
\langle \Omega \vert \hat c^{\dagger}_{i\sigma} \hat c_{j\sigma} \vert x_{\mathbf{s}} \rangle
&=
g_{ij,\sigma}(x_{\mathbf{s}})
\end{align}
and the hopping part ($\hat H_t$) of~Eq.\eqref{terms2} can be computed from:
\begin{align}
\langle \Omega \vert 
\hat c_{i\sigma} 
\hat c^{\dagger}_{k\sigma^\prime} 
\hat c_{l\sigma^\prime} 
\hat c^{\dagger}_{j\sigma} 
\vert x_{\mathbf{s}} \rangle
=&
-g_{jikl,\sigma\sigma^\prime}(x_{\mathbf{s}})  
+ \delta_{ij} g_{kl,\sigma^\prime}(x_{\mathbf{s}})
\nonumber
\\
&+ \delta_{jl} \delta_{\sigma \sigma^\prime} ( \delta_{ik} - g_{ki,\sigma}(x_{\mathbf{s}}))
\\
\langle \Omega \vert 
\hat c^{\dagger}_{i\sigma} 
\hat c^{\dagger}_{k\sigma^\prime} 
\hat c_{l\sigma^\prime} 
\hat c_{j\sigma} 
\vert x_{\mathbf{s}} \rangle
=&
g_{ijkl,\sigma\sigma^\prime}(x_{\mathbf{s}})  
-  \delta_{jk} \delta_{\sigma \sigma^\prime}
g_{il,\sigma}(x_{\mathbf{s}}).
\end{align}
This minimizes the number of Pfaffian calculations. It in principle generates terms for every  $0 \le i,j < N $, but it is convenient to impose the translational invariance at every Monte Carlo sample if it is satisfied, to reduce the memory cost and Monte Carlo noise. If so, only the terms $i=0$ and $0 \le j-i < N $ are required to compute. Finally, it is important to impose hermicity at every sampling too, so that:
\begin{align}
\langle \psi_{im} \vert \hat c_{i\sigma} \hat c^{\dagger}_{j\sigma} \vert \psi_{jn} \rangle 
= &  
\Big(\langle \psi_{im} \vert \hat c_{i\sigma} \hat c^{\dagger}_{j\sigma} \vert \psi_{jn} \rangle 
\\
&+
\langle \psi_{jn} \vert \hat c_{j\sigma} \hat c^{\dagger}_{i\sigma} \vert \psi_{im} \rangle^\star
\Big)/
{2}
\nonumber
\\
\langle \psi_{im} \vert \hat c_{i\sigma} \hat H \hat c^{\dagger}_{j\sigma} \vert \psi_{jn} \rangle
=&
\Big(
\langle \psi_{im} \vert \hat c_{i\sigma} \hat H \hat c^{\dagger}_{j\sigma} \vert \psi_{jn} \rangle
\\
&+
\langle \psi_{jn} \vert \hat c_{j\sigma} \hat H \hat c^{\dagger}_{i\sigma} \vert \psi_{im} \rangle^\star
\Big)/{2}
\nonumber
\end{align}
and so on. These two optimizations greatly reduce noise in the resulting data.


\section{Dependence of accuracy on the level and number of  included excitations}

\label{decomposition_appendix}

In this Appendix, we show effect of increasing
the range $T$ for $\delta$ in Eq.~\eqref{charge_exc}.

In Figs.~\ref{chain64_decomp_mott} and~\ref{chain64_decomp}, we show the dependence on the number of excitations for the result present in Fig.~\ref{chain16_U8_Ne16}(c) and~\ref{chain16_U8_Ne14}(c) respectively.
We see that the first few neighbors are crucial to get better results. 
Between $\delta_{max} = 6$ and $\delta_{max} = 7$, the result does not change much even if we add $83$ new excitations suggesting the overall convergence to the exact results aside from detailed thin structures near the Fermi level in Fig.~\ref{chain64_decomp}.


\begin{figure}
\centering
\includegraphics[width=0.9\linewidth]{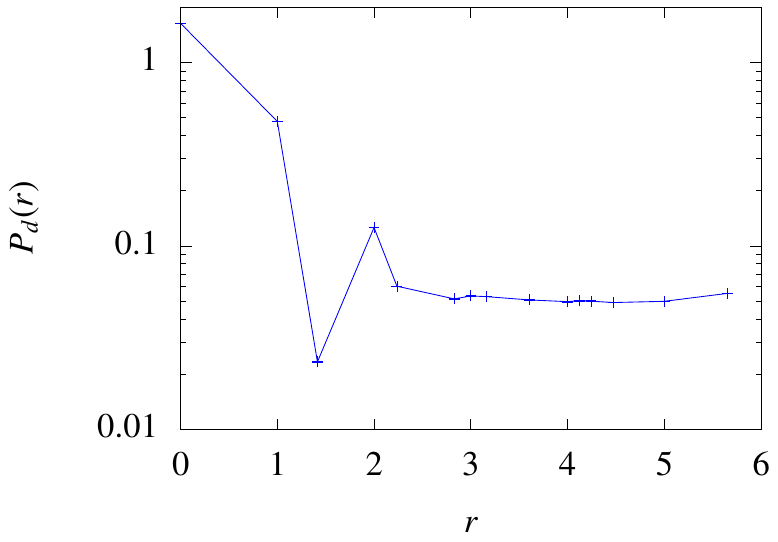}
\caption{Superconducting correlation function $P_d(r)$ as a function of pair-pair distance $r$. The cross data points also contain (very small) error bars.}
\label{Pd_r}
\end{figure}

\section{Superconducting ground state}

\label{Pd}

The solution presented in Fig.~\ref{8x8_Ne56_Akw}(c) has a superconducting order. This can be seen from the gap seen in the spectral weights plotted in both Fig.~\ref{8x8_Ne56_Akw}(c) and Fig.~\ref{fig:gap}. This can be seen more directly by measuring the $d$-wave superconducting correlation function 
\begin{align}
P_{d}(\boldsymbol{r})=\frac{1}{2 N_{s}} \sum_{i}\left\langle\Delta_{d}^{\dagger}(\boldsymbol{r}_i) \Delta_{d}\left(\boldsymbol{r}_i+\boldsymbol{r}\right)+\Delta_{d}(\boldsymbol{r}_i) \Delta_{d}^{\dagger}\left(\boldsymbol{r}_i+\boldsymbol{r}\right)\right\rangle
\end{align}
with 
\begin{align}
\Delta_{d}\left(\boldsymbol{r}_{i}\right)=\frac{1}{\sqrt{2}} \sum_{r} g(\boldsymbol{r})\left(c_{r_{i} \uparrow} c_{r_{i}+r \downarrow}-c_{r_{i} \downarrow} c_{r_{i}+r \uparrow}\right)
\label{Delta}
\end{align} 
with the $d$-wave form factor $g$ defined as 
\begin{align}
g(r)=\delta_{r_x ,0}(\delta_{r_y,1}+\delta_{r_y,-1})-
\delta_{r_y,0}(\delta_{r_x,1}+\delta_{r_x,-1})
\label{g-formfactor}
\end{align} 
in~\cite{ido_competition_2018}. This function is shown in Fig.~\ref{Pd_r}. The long-range superconducting order is indicated by the saturation to a nonzero value around $\sim0.055$ at long distance.


%

\end{document}